# Laser guided ionic wind

Shengzhe Du[1, §], Tie-Jun Wang[1,*, §], Zhongbin Zhu[1], Yaoxiang Liu[1], Na Chen[1], Jianhao Zhang[1], Hao Guo[1], Haiyi Sun[1], Jingjing Ju[1], Cheng Wang[1], Jiansheng Liu[1, #], See Leang Chin[2], Ruxin Li[1, †]  and Zhizhan Xu[1,‡]

[1] *State Key Laboratory of High Field Laser Physics, Shanghai Institute of Optics and Fine Mechanics, Chinese Academy of Sciences, China*
[2] *Centre d'Optique, Photonique et Laser (COPL) and Département de physique, de génie physique et d'optique, Université Laval, Québec, Québec G1V 0A6, Canada*

*tiejunwang@siom.ac.cn,
#michaeljs_liu@mail.siom.ac.cn,
†ruxinli@mail.shcnc.ac.cn
‡zzxu@ mail.shcnc.ac.cn
§ these two contributed equally to the work

## Abstract

We report on a method to experimentally generate ionic wind by coupling an external high voltage electric field with an intense femtosecond laser induced air plasma filament. The measured ionic wind velocity could be as strong as >4 m/s. It could be optimized by changing the applied electric field and the laser induced plasma channel. The experimental observation was qualitatively confirmed by a numerical simulation of spatial distribution of the electric field. This technique is robust and free from sharp metallic electrodes for coronas; it opens a way to optically generate ionic wind at a distance.

## 1. Introduction

Ionic wind, also called coronal wind or electric wind [1, 2], is an air flow generally driven by the electric field created by applying a high voltage to an electrode system. Initially the electrons in the ambient air resulting from ionization of air molecules by cosmic rays are accelerated under the high electric field near the electrode and undergo inelastic collisions. Consequently, more molecules are ionized in the process of electron avalanche ionization. The ions resulting from the ionization are also accelerated by the electric field in the opposite direction of the motion of the electrons and transfer their momenta to air particles via collisions, initiating a drag of the bulk air which is referred to as the ionic wind. Ionic wind has raised great interests and has found applications [3-9]. Due to their robustness, simplicity, low power consumption, and ability for real-time control at high frequency, plasma based actuators have been widely used in aerodynamic applications [3]. As a successful demonstration, Deep Space 1, the first mission of NASA's New Millennium Program propelled by ion thruster engine, was launched on October 24, 1998 [4, 5]. An ionic wind generator has been suggested for a next-generation cooling device for LEDs and other electronic devices because of its high cooling performance, light and compact size, low noise, immune from vibration etc. [6]. The mobilities of ions from ionic wind created by corona discharges has also been used for gas diagnostics, for instance to detect and measure gas contaminants through the mobility spectrum of the ions [7]. In addition, ionic wind could induce precipitations such as rain and snow formation in a cloud chamber [8, 9]. Traditionally, an ionic wind generator strictly relies upon the configuration design of electrodes, such as needle-to-plate and needle-to-cylinder types [3-9]. Those setups are mostly limited in different kinds of restricted space or fixed location due to the fixed mechanical design of the electrodes.

In this work we report on a brand new method to generate ionic wind by coupling a high voltage electric field with an intense femtosecond laser ionized air plasma channel, namely filament. High voltage electric field was efficiently guided along the laser ionized air plasma channel resulting in the coronas at the end of the channel for ionic wind generation. This method is robust and immune to the specific design of traditional ionic wind generator, which also has potential application at a distance, even remotely.

## 2. Experimental setup

A schematic of the experimental setup is shown in Fig. 1. Experiments were conducted using a Ti:sapphire chirped pulse amplification (CPA) laser system producing pulse energy of up to 6.85 mJ with central wavelength of 800 nm at a repetition of 1 kHz. The full-width half-maximum (FWHM) length of each pulse is 25 fs. The laser beam was focused by a convex lens with a focal length of 50 cm to create a stable filament into a home-made Faraday cage. A spherical copper electrode with a special structure was designed to efficiently couple a high voltage electric field to the filament [10]. The main part of the electrode is a copper ball with a diameter of approximately 40 mm and there is a cylindrical channel punched through the ball as shown as the inset of Fig. 1. The diameter of the cylinder is 4 mm which is wide enough for the filaments to get through. The spherical copper electrode was connected to a DC high voltage power supply with a maximal output of positive 100 kV/1000 W.

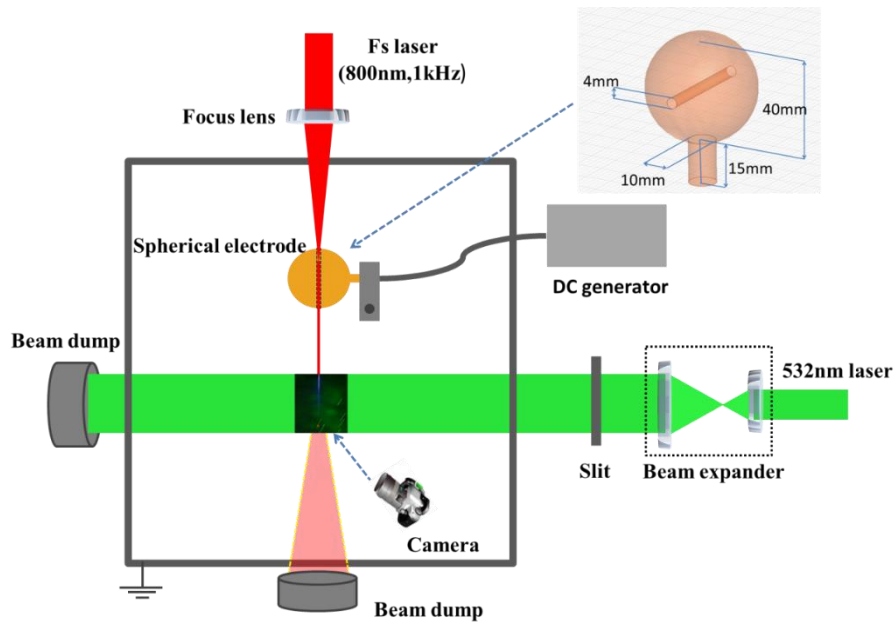

Fig.1 Experimental setup for ionic wind generation by laser guided coronas. The spherical electrode was held with an insulated plastic holder. The electrode design is shown in the inset. A good coupling was achieved by sending laser filament through the cylindrical channel which was punched through the spherical electrode.

Two methods were adapted to monitor the ionic wind velocity. One is based on the imaging of moving particles as shown in Fig. 1. Coarse particles ($CaCO_3$/$CaSO_4$ powders) were sprayed by hand into the Faraday cage. The diameters of these coarse particles are mostly 20~100 μm. A continuous wave (CW) 532-nm laser beam with 3.0 W output power after being expanded in diameter and truncated by a 30 mm (length) × 6 mm (width) slit was used to shine onto the particles. The slice of the laser beam passed through the horizontal plane where the 800 nm Ti: sapphire laser pulses propagated at 90 deg. as shown in Fig. 1. By recording the scattered 532 nm laser from moving particles with a digital camera of 23 frames per second (Nikon D7200), ionic wind velocity in different regions was estimated by simply doing divisions between the distances of particle trajectories and the exposure times, S, they took. Note that the actual ionic wind velocity should be higher than the measured value, owing to a larger average mass of the coarse particles than the air molecules. The second method was based on a hot wire anemometer probe (HHF-SD1). The probe was placed 15 cm away from the ionization region around the tip of the filaments to directly measure the ionic wind in a relatively far and safe area.

## 3. Results and Discussion

### 3.1 Applied voltage dependence.

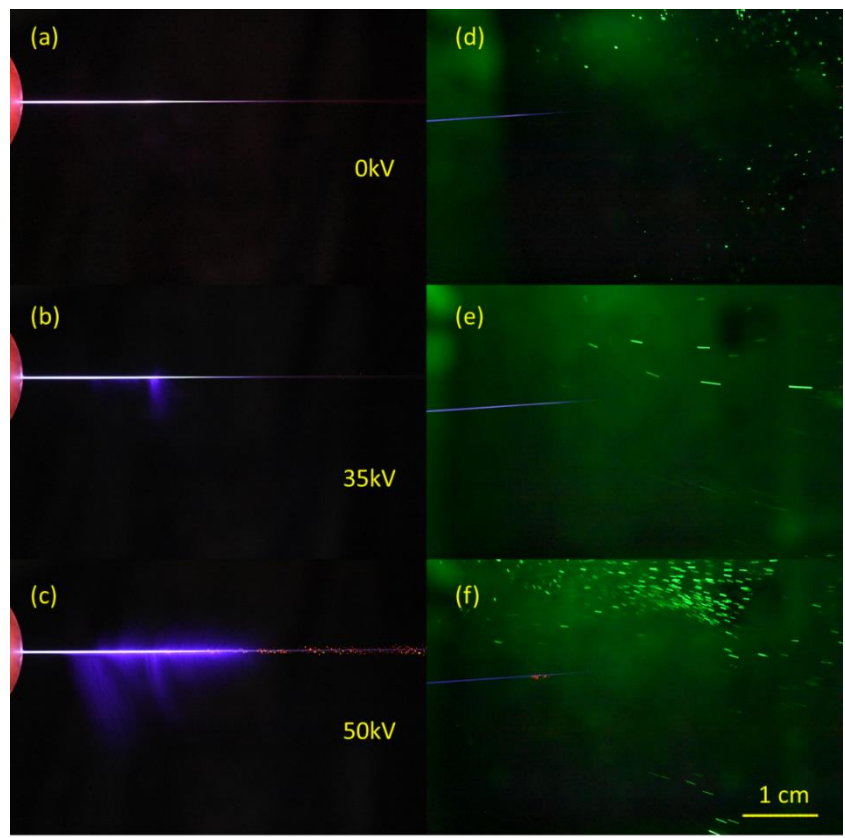

Fig. 2 Real-color images of coronas generated along the filament in air with applied voltage at (a) 0 kV; (b) 35 kV; (c) 50 kV, and the images of the particle movements at the same voltages ((d)-(f)). The recording parameters of the camera for (d)-(f) are S = 1/50 s, F = 4, and ISO = 25600 (d); S = 1/200 s, F = 4, and ISO = 25600 (e); S = 1/500 s, F = 4, and ISO = 25600 (f), respectively.

The filament passed through the cylindrical hole of the electrode (Fig. 2(a)). The applied voltage was tuned from 0 kV to 50 kV. Leader and streamer types of corona discharges occurred around the tip and the length of the filaments (Fig. 2(b) and Fig. 2(c)) [11]. At low applied voltage of <30 kV, only leader type of coronas along the laser filament was observed. When the voltage was further increased, streamer type of coronas along the laser filament occurred and became much more significant (Fig. 2(c)). Note that the newly generated streamers of the corona in Fig. 2(b) and Fig. 2(c) are not symmetrical along the filament because of the asymmetry of the copper electrode. After spraying coarse particles into the cage, asymmetric particle trajectories around the filament driven by ionic wind were clearly observed and recorded by the camera (Fig. 2(d)-2(f)). A short video of the movement in Fig. 2(f) can be seen here.

From the recorded pictures of particles' movements (e.g. Fig. 2(d)-2(f)), maximum velocities of these coarse particles were calculated under different high voltages (Fig. 3). The computed rectangular (9.99 cm × 6.65 cm) area in Fig. 2 was set near the front tip of the filament where maximum velocities were observed. The background particles (applied voltage = 0 kV) slowly floated around with a velocity of ~0.05 m/s.

When high voltage electric field was coupled onto the air plasma filament, there were two regimes of maximum velocity of air flow when the applied voltage was increased as shown in Fig. 3: in the first regime from 0 kV to ~30 kV, the velocity was linearly proportional to the voltage; the velocity was also linearly proportional to the voltage in the second regime when the voltage was above 35 kV, but with a larger slope. This behavior is different from the trend of the ionic wind induced by a metallic electrode, in which only one linear dependence was reported [12]. A maximum velocity of >4 m/s was measured at the voltage of 50 kV.

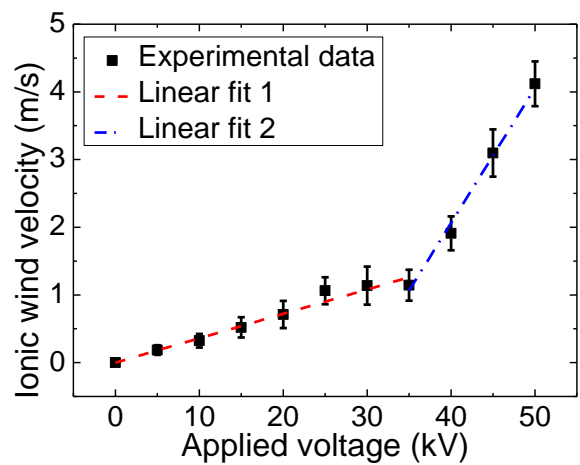

Fig. 3 The maximum ionic wind velocity as a function of the applied voltage. Squared points are the measurement results and the red dashed and blue dashed dot line are the linear fits.

## 3.2 Air plasma length dependence

By fixing the laser pulse at 1 kHz/25 fs/6.85 mJ and the applied voltage at 30 kV where mostly leader type of corona discharges was observed at the filament tip, the dependence of the ionic wind velocity on the air plasma filament length (which was defined as the length outside the electrode) was investigated by moving the filament through the electrode. As shown in Fig. 4(a), when the external filament length was tuned from 3.61 cm to 7.35 cm, the resultant wind velocity was almost constant. No significant change was observed. Since the length of the plasma filament could be controlled by playing with laser parameters, such as energy, chirp, numerical aperture etc., from sub-meter to several tens of meters [13, 14], the air plasma guided ionic wind can be potentially generated at a far distance, even remotely.

In another experiment, the volume and the plasma density of the filament were changed by varying the laser pulse energy. The high voltage was fixed at 30 kV. When the laser energy was tuned from 2.85 mJ to 6.85 mJ, the maximum ionic wind velocity varied from 0.57 m/s up to 1.15 m/s (Fig. 4(b)). Two times increase in ionic wind was achieved. This result indicates that the air plasma channel induced ionic wind could be enhanced by improving the pulse energy.

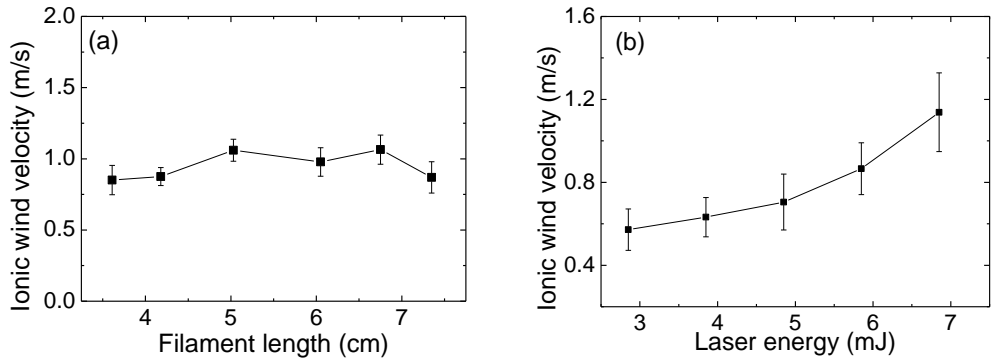

Fig. 4 (a) measured ionic wind velocity as a function of (a) air plasma length outside the electrode and (b) the laser pulse energy. See more details in the text.

**3.3 Angular distribution of the ionic wind**

By using a hot wire anemometer probe (HHF-SD1), the ionic wind velocity was measured at a radial distance of 15 cm from the tip of the laser filament. We note that the hot wire anemometer is a thin metallic wire based sensor, which cannot work at the corona region because of the strong electric field. We have to put it a little far away from the coronas although it is very accurate. The detection schematic is shown in Fig. 5(a). Laser propagation direction is defined as angle of 0 degree. The angle in the clockwise direction is positive. 1 kHz/25 fs/6.85 mJ femtosecond laser pulses were used to create plasma filaments with a 50 cm focal length lens. The high voltages were fixed at 10 kV, 20 kV and 30 kV, respectively. To reduce the influence of high voltage electric field on the anemometer, an electrostatic shielding to the probe was done by twining grounded metallic mesh on the surface. As shown in Fig. 5(b), peak velocity occurs at 0 deg., which is the direction of laser propagation. The wind velocity is stronger where stronger coronas occurred at higher applied voltage.

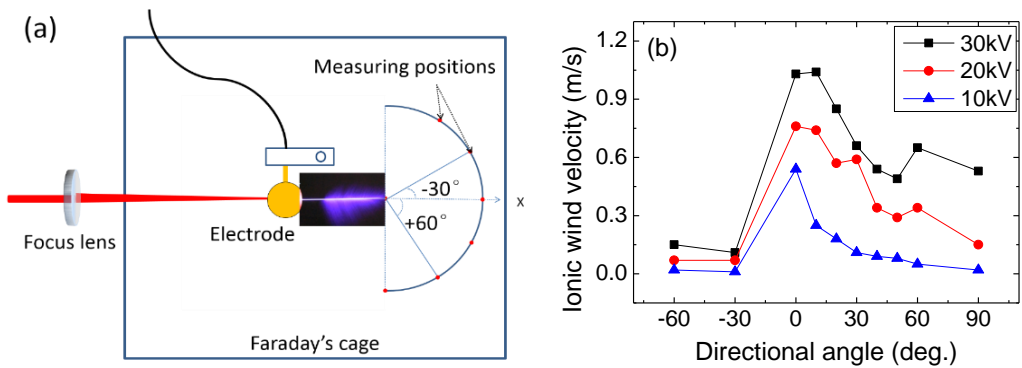

Fig. 5 (a) the schematic of directional angle measurement of ionic wind. The electrode was held by an insulated plastic holder as shown in the figure. (b) the measured maximum ionic wind velocity as a function of detection angle. 1 kHz/25 fs/6.85 mJ femtosecond laser pulse was used to create plasma filament with a 50 cm focal length lens. The high voltage was fixed at 10 kV, 20 kV and 30 kV, respectively.

**3.4 Numerical consideration and discussions**

In order to have a spatial distribution of the electric field **E** around the air plasma filament, the following Poisson equation and Maxwell equation were solved through finite element analysis:

$$\nabla \cdot (\varepsilon_r \cdot \varepsilon_0 \nabla \Phi) = -\rho_V \quad (1)$$

$$E = -\nabla \Phi \quad (2)$$

where $\varepsilon_r$ is the relative dielectric constant, $\varepsilon_0$ the permittivity of vacuum, $\Phi$ the electric scalar potential, and $\rho_V$ the density of volume charges. Under a fixed applied voltage at 50 kV, the distributions of electric field intensity **E** around the spherical electrode (radius of curvature 20 mm) with and without an artificial filament are presented in Fig. 6. In the simulation, the artificial filament was regarded as a uniform plasma cylinder with a length of 6.75 cm (the radius of curvature r of the plasma channel was set at 0.05 mm and the electrical conductivity of the channel (filament) was assumed 100 siemens/s according to ref. 15.

When applying the high voltage of 50 kV to the spherical electrode only, the maximum electric field strength on the surface of the electrode in Fig. 6(a) is about $2 \times 10^6$ V/m which is below the electric field threshold of $(E_r)_{sphere} = 4.05 \times 10^6$ V/m for a streamer corona to occur, according to ref. 16. This means there won't be any streamer coronas with the spherical electrode alone. When the laser filament was employed to couple with the high electric field in the current configuration (Fig. 6 (b)), the resultant electric field along the artificial laser filament is ~$6.2 \times 10^7$ V/m which is well above the threshold ($(E_r)_{fila} = 1.66 \times 10^7$ V/m) for streamer coronas generation according to ref. 17. Note that the calculated electric field threshold for streamer generation with filament ($(E_r)_{fila}$) is higher than that of spherical electrode ($(E_r)_{sphere}$) under the same applied voltage. This may be due to the low conductivity of filament as compared to metallic wire although filament has sharp ends. Strong electric field around the tip of filament leads to coronas generation through avalanche ionization. Indeed, no streamer coronas were observed in the absence of plasma filament in the experiment at 50 kV. But clear streamer coronas were generated and observed along the plasma filament when the laser filament was added (Fig. 2(c)). These coronas are the sources of ionic wind. The experimental observation agrees well with the numerical predication. Therefore, the ionic wind in our experiment only originated from the laser guided coronas, not from the spherical electrode. In Fig. 6(b), there is weak field regime (near zero field) not far away from the exit of the electrode hole. This is due to a balance between the electric field from the electrode and the filament guided electric field. The jitter of the pulses would smear out the near zero point because of its high repetition rate of 1 kHz. Even so, one can see this effect in the experiment; i.e. for a short distance outside the hole, there is no streamer at very high voltage (Fig. 2(c)) while after this short distance, streamers come out almost continuously along the filament till its end tip.

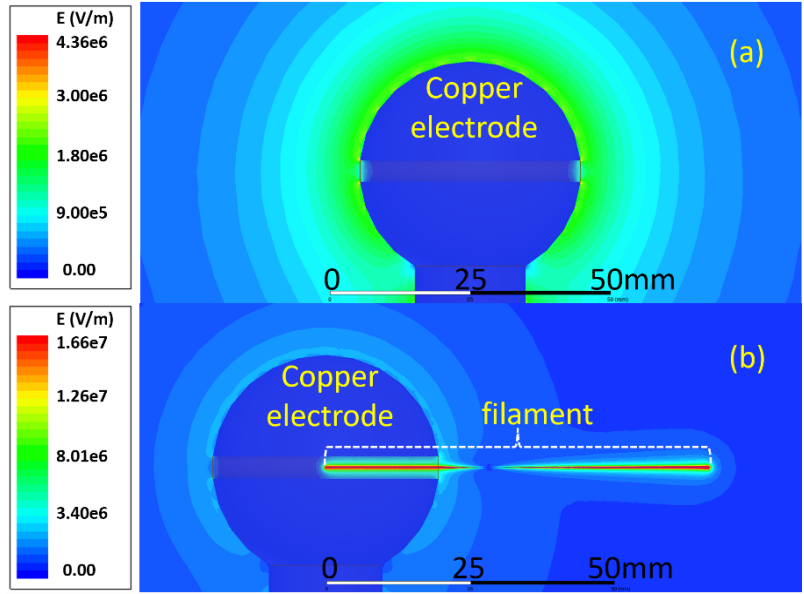

Fig. 6 Electric field distribution around the spherical electrode when a high voltage of 50 kV was used (a) without filament and (b) with filament.

### 3.5. Discussion

The air plasma filament [18-21] created by femtosecond laser pulses through multiphoton/tunneling ionization can serve as a long conductor like a conducting wire offering early free electrons. When applying positive high voltage to the spherical copper electrode, the free electrons inside the external part of the filament would be accelerated towards the electrode and positive ions are pushed towards the tip of the filament. In the early stage of increasing the applied voltage (<30 kV), the weak electric field at the tip of the filament resulted in a weak avalanche ionization for leader type of corona generation, which may be responsible for the weak ionic wind generation in the first linear regime in Fig. 3. As the applied voltage keeps increasing, the positive ions at the sharp end of the filament produce much higher positive electric field, as a consequence, significantly stronger leaders and streamers are generated leading to much stronger ionic wind generation at the high voltage in the second linear regime of Fig. 3. The ionic wind originates from the coronas from the filaments (Fig. 5); hence, it can be generated at a distance (Fig. 4(a)) and optimized by shaping the laser induce plasma volume (Fig. 4(b)) and the external electric field [10].

## 4. Conclusion

We experimentally demonstrated that a strong ionic wind (>4 m/s) can be generated by coupling an external high voltage electric field with a femtosecond laser-induced air plasma filament. The ionic wind relies upon the coronas along the plasma filament when an external high electric filed is applied. The dependence of the ionic wind velocity on the applied high voltages, plasma channel and laser characteristics have been systematically investigated. The experimental results are qualitatively understood with a numerical simulation. The approach reported in the work paves a way to optically generate ionic wind at a distance, even remotely.

**Acknowledgments**

This work was supported in part by the Strategic Priority Research Program of the Chinese Academy of Sciences (Grant No. XDB16000000), Key Project from Bureau of International Cooperation Chinese Academy of Sciences (Grant No. 181231KYSB20160045) and 100 Talents Program of Chinese Academy of Sciences, China. SLC acknowledges the support from Laval University in Canada.

**References**

[1] Myron Robinson, "A history of the electric wind", American Journal of Physics, 30, 366-372 (1962).
[2] M. Goldman, A. Goldman, and R. S. Sigmond, "The corona discharge, its properties and specific uses", Pure & Applied Chemistry 57, 1353-1362 (1985).
[3] Eric Moreau, "Airflow control by non-thermal plasma actuators", Journal of Physics D: Applied Physics 40, 605-636 (2007).
[4] Marc D. Rayman, and David H. Lehman, "Deep Space One: NASA's first deep-space technology validation mission", Acta Astronautica 41, 289-299 (1997).
[5] Marc D. Rayman, "The Deep Space 1 extended mission: challenges in preparing for an encounter with comet Borrelly", Acta Astronautica 51, 507-516 (2002).
[6] Dong Ho Shin, Joon Shik Yoon, and Han Seo Ko, "Experimental optimization of ion wind generaor with needle to parallel plates for cooling device", International Journal of Heat and Mass Transfer 84, 35-45 (2015).
[7] A. Goldman, R. Haug, and R. V. Latham, "A repulsive-field technique for obtaining the mobility-spectra of the ion species created in a corona discharge", Journal of Applied Physics 47, 2418-2423 (1976).
[8] G. S. P. Castle, "Industrial applications of electrostatics: the past, present and future", Journal of Electrostatics, 51-52, 1-7 (2001).
[9] Jingjing Ju, Tie-Jun Wang, Ruxin Li, Shengzhe Du, Haiyi Sun, Yonghong Liu, Ye Tian, Yafeng Bai, Yaoxiang Liu, Na Chen, Jingwei Wang, Cheng Wang, Jiansheng Liu, S. L. Chin, and Zhizhan Xu, "Snowfall induced by corona discharge", arXiv:1607/05125[physics.gen-ph].
[10] Shengzhe Du, Zhongbin Zhu, Yaoxiang Liu, Tie-Jun Wang, and Ruxin Li, "Optimization design scheme of femtosecond laser induced corona discharge", Chinese Journal of Lasers 44, 601009 (2017). (in Chinese)
[11] Tie-Jun Wang, Yingxia Wei, Yaoxiang Liu, Na Chen, Yonghong Liu, Jingjing Ju, Haiyi Sun, Cheng Wang, Haihe Lu, Jiansheng Liu, See Leang Chin, Ruxin Li, and Zhizhan Xu, "Direct observation of laser guided corona discharges", Scientific Reports 5, 18681 (2015).
[12] Longnan Li, Seung Jun Lee, Wonjung Kim, and Daejoong Kim, "An empirical model for ionic wind generation by a needle-to-cylinder dc corona discharge", Journal of Electrostatics 73, 125-130 (2015).
[13] S. L. Chin, *Femtosecond Laser Filamentation* (Springer, New York, 2010).
[14] J. Kasparian, J.-P. Wolf, "Physics and applications of atmospheric nonlinear optics and filamentation", Opt. Express **16**, 466-493 (2008).
[15] S. Tzortzakis, M. A. Franco, Y.-B. Andre, A. Chiron, B. Lamouroux, B. S. Prade, and A.

Mysyrowicz, "Formation of a conducting channel in air by self-guided femtosecond laser pulses", Physical Review E 60, R3505-R3507 (1999).

[16] G. Hartmann, "Theoretical evaluation of Peek's law", IEEE Transactions on Industry Applications IA-20, 1647-1651 (1984).

[17] F. W. Peek, Dielectric phenomena in high voltage engineering (McGraw-Hill, 1929).

[18] S. L. Chin, S. A. Hosseini, W. Liu, Q. Luo, F. Théberge, N. Aközbek, A. Becker, V. P. Kandidov, O. G. Kosareva, and H. Schroeder, "The propagation of powerful femtosecond laser pulses in optical media: physics, applications, and new challenges", Canadian Journal of Physics **83**, 863-905 (2005).

[19] A. Couairon and A. Mysyrowicz, "Femtosecond filamentation in transparent media", Phys. Rep. **441**, 47-189 (2007).

[20] L. Bergé, S. Skupin, R. Nuter, J. Kasparian, J.-P. Wolf, "Ultrashort filaments of light in weakly ionized, optically transparent media", Rep. Prog. Phys. **70**, 1633-1713 (2007).

[21] S. L. Chin, T.-J. Wang, C. Marceau, J. Wu, J. S. Liu, O. Kosareva, N. Panov, Y. P. Chen, J.-F. Daigle, S. Yuan, A. Azarm, W. W. Liu, T. Saideman, H. P. Zeng, M. Richardson, R. Li, and Z. Z. Xu, "Advances in intense femtosecond laser filamentation in air", Laser Phys. **22**, 1-53 (2012).

**Supplementary video: Movie S1**

Movement of laser guided ionic wind recorded a digital camera. Laser plasma channel was created by focusing 6.8 mJ, 30 fs laser pulse with 40 cm focal length lens. DC high voltage of 20 kV was applied onto the channel through the designed electrode.